Introducing Nonuniform Strain to Graphene Using Dielectric Nanopillars

Hikari Tomori[1,2], Akinobu Kanda[1,2] [*], Hidenori Goto[1,2], Youiti Ootuka[1], Kazuhito Tsukagoshi[2,3], Satoshi Moriyama[3], Eiichiro Watanabe[4], and Daiju Tsuya[4]

[1]Institute of Physics and Tsukuba Research Center for Interdisciplinary Materials Science (TIMS), University of Tsukuba, Tsukuba, Ibaraki 305-8571, Japan

[2]CREST, Japan Science and Technology Agency, Kawaguchi, Saitama 332-0012, Japan

[3]International Center for Materials Nanoarchitectonics (MANA), National Institute for Materials Science (NIMS), Tsukuba, Ibaraki 305-0044, Japan

[4]Nanotechnology Innovation Center, NIMS, Tsukuba, Ibaraki 305-0047, Japan

A method for inducing nonuniform strain in graphene films is developed. Pillars made of a dielectric material (electron beam resist) are placed between graphene and the substrate, and graphene sections between pillars are attached to the substrate. The strength and spatial pattern of the strain can be controlled by the size and separation of the pillars. Application of strain is confirmed by Raman spectroscopy as well as from scanning electron microscopy (SEM) images. From SEM images, the maximum stretch of the graphene film reaches about 20%. This technique can be applied to the formation of band gaps in graphene.

[*] E-mail address: kanda@lt.px.tsukuba.ac.jp

Graphene, a two-dimensional honeycomb lattice made out of a single sheet of carbon atoms, attracts much attention not only from the viewpoint of its Dirac fermionic fundamental properties but also for its possible applications. Graphene is inherently a zero-gap semiconductor with the conical band structure, in which the conduction and valence bands touch each other at the Dirac point.[1] Thus, for the successful application of graphene to electronic devices, especially to field-effect transistors (FETs), the formation of a large band gap is indispensable. Several methods have been proposed for opening band gaps in graphene, such as the formation of graphene nanoribbons,[2] the application of perpendicular electric fields to bilayer graphene,[3] and the chemical modification of graphene.[4,5] However, experimental trials to confirm band gaps performed so far showed a strong contribution of variable-range-hopping conduction in the band gap, causing a significant leakage current.[5-7] This means that the presence of localized states in the band gap due to disorder (edge roughness, impurities, lattice defects, and so on) significantly deteriorates the performance of graphene FETs. There is a pressing need to enhance the band gap and improve its quality in graphene.

Recently, another method to induce a band gap was proposed, which was based on the generation of a pseudo-magnetic field by applying a shear strain to graphene;[8] Owing to the chiral character of the charge carriers in graphene, it was predicted that a designed strain along three main crystallographic directions induces strong gauge fields, resulting in almost uniform magnetic fields. A shear strain of 10% applied to a 100 nm graphene nanoflake induces a uniform magnetic field of ~ 20 T near the center of the flake, and the magnitude of the pseudomagnetic field scales as $\Delta_m/L$, where $\Delta_m$ is the maximum strain and $L$ is the flake size. Besides, when a strain with triangular symmetry (triangular strained superlattice) is introduced to bulk graphene, e.g., by placing a graphene sheet on top of a corrugated surface with a triangular landscape, the resulting periodic pseudomagnetic field with triangular symmetry leads to a Landau-level-like energy spectrum with multiple energy gaps at high energies. Although the numerically obtained energy gap of ~ 10 meV is relatively small for a strain $\Delta_m$ ~ 0.1% with period $L = 40$ nm,[8] the gap can be enhanced by adjusting the strain and the period. Thus, the strain-engineered band gap is a possible breakthrough for graphene-based electronics.

Experimentally, strain-induced Landau levels were observed in a scanning tunneling microscopy study on the top of nanobubbles which were naturally formed in chemical-vapor-deposition (CVD)-grown graphene.[9] The corresponding pseudo-magnetic field exceeded 300 T. On the other hand, designed pseudo-magnetic fields or strain-induced band gaps in graphene have not been attained yet, mainly because a practical method for introducing a designed nonuniform strain is lacking. Here, we note that neither uniaxial nor isotropic strain produces any pseudomagnetic field.[8]

In this letter, we describe a technique to introduce a designed nonuniform strain to graphene. For this purpose, we place nanopillars made of a dielectric material (electron beam resist) between a graphene film and the Si substrate. The strength and spatial pattern of the strain can be designed by adjusting the position and size of the pillars.

For the fabrication of nanopillars on the surface of Si substrates, we take advantage of an uncommon property of e-beam resists; excess exposure to e-beam makes resists such as lift-off resist (LOR) and polymethyl methacrylate (PMMA) insoluble even in their developers and removers. The fabrication process of nanopillars is illustrated in Fig. 1. First, a graphene film is placed on a Si substrate coated with

LOR-1A resist (MicroChem) (Fig. 1(a)). One can fabricate additional electrodes at this stage for electrical measurement by e-beam lithography, metal deposition, and lift-off, using another resist, e.g., PMMA, which uses developers and removers different from those for LOR. Then, the pattern for pillars is transferred to the LOR resist with an e-beam dose of 60,000 $\mu C/cm^2$, which is more than 100 times larger than the normal exposure (Fig. 1 (b)). Next, we remove the LOR resist without the excess e-beam exposure by dipping the substrate in the remover, N-methyl-2-pyrrolidone, for 5 min, rinse it with 2-propanol for 30 sec, and then put the substrate into distilled water for 30 sec. Finally, we dry the substrate by nitrogen blow. If we skip the dip into distilled water, the graphene film is suspended over the LOR pillars as shown in Figs. 1(c), 1(e), and 1(f). On the other hand, a graphene film after dipping into distilled water is attached to the substrate, as illustrated in Fig. 1(d), indicating that the graphene film is nonuniformly strained. Throughout this study, the graphene films were prepared by mechanical exfoliation of kish graphite,[10] and the thickness of the films was checked by the contrast of optical images[10] and Raman spectroscopy.

To estimate the amount of stretch in graphene films, we place additional PMMA disks on top of the LOR pillars. Here, after depositing graphene (Fig. 1(a)), we coat the substrate with thin PMMA, and then expose it to excess e-beam (Fig. 1(b)). Figure 2 shows scanning electron microscopy (SEM) images of such samples. In Fig. 2(a), each PMMA disk is exactly on top of the corresponding LOR pillar, indicating that the graphene did not shift laterally when stretching, as illustrated in Figs. 2(b) and 2(c). The average stretch is defined as the increase of the length of a graphene section between a pair of pillars after the stretch divided by the original graphene length (separation of the pillars). From Fig. 2(a), one can estimate the average stretch of graphene between pillars, which ranges from 6 to 20%, depending on the direction, as shown in Fig. 2(d), in which pillars A to D correspond to those in Fig. 2(a). The absence of the triangular symmetry in the average stretch is presumably because of a minimal lateral shift of graphene, which is undetectable by the present method. This shift also causes wrinkles in Fig. 2(a). Figure 2(e) shows an another example, in which the top PMMA disks are shifted in the direction perpendicular to the graphene edge (indicated by the dashed line). The amount of the shift becomes larger as pillars approach the graphene edge. Even in this case, though, the graphene film stretches along the shift direction with an average stretch of 7%, as estimated from the SEM image.

To confirm the nonuniform strain in graphene, we performed spatially resolved micro-Raman spectroscopy. The Raman spectra were acquired using a laser excitation of 532 nm (2.33 eV) with an incident power of 100 μW and a spot size of ~ 0.5 μm. The result is shown in Fig. 3. Figure 3(a) is an SEM image of the sample, in which a triangular lattice of LOR pillars 130 nm high and 200 nm wide is placed underneath graphene. The separation of pillars is 1.5 μm. The average stretch of graphene between adjacent pillars is estimated to be ~ 6 - 9% from SEM images., depending on the direction. Figure 3(b) shows the spatial distribution of the Raman signal in the area indicated by a yellow square in Fig. 3(a). Here, the average signal intensity between 2665 and 2675 cm$^{-1}$ is plotted in a gray scale. White (black) regions mean high (low) Raman intensity. It is clear that the spatial variation of the Raman signal corresponds to the triangular lattice of pillars, indicated by red circles in Fig. 3(b). To see this in more detail, we look at the Raman spectrum at several points, $A_1$, $A_2$, $B_1$, …, $E_2$, indicated in Fig. 3(c). Here, $A_i$ ($i = 1, 2$) is on a pillar and $B_i$ to $D_i$ are located near the midpoint of adjacent pillars. Points $X_1$ and $X_2$ (X = A, ..., D) are equivalent in the triangular lattice.

$E_1$ and $E_2$ are near the graphene edge. The Raman spectra of these points around the 2D band are shown in Figs. 3(d) - 3(f). The 2D band comes from a second-order double resonant process of zone-boundary phonons. The full width at half maximum (FWHM) of ~ 30 cm$^{-1}$ is the fingerprint of single layer graphene (SLG).[11] On the pillars (Fig. 3(d)), the 2D peak is located at $(2663 \pm 1)$ cm$^{-1}$, which downshifts in comparison with the value of ~ 2680 cm$^{-1}$ for graphene placed on a Si substrate (without pillars),[11,12] as shown in Fig. 3(g). Note that the LOR pillars (without graphene) do not show any Raman peak around the 2D band. On the other hand, near the midpoint between adjacent pillars (Fig. 3(e)), the spectrum is almost independent of the direction within the accuracy of the measurement, and exhibits a maximum at $(2670 \pm 2)$ cm$^{-1}$. The difference of the 2D peak positions in Figs. 3(d), 3(e), and 3(g) presumably originates from the variation of strain. Actually, near a graphene edge, where one can expect a relaxation of strain due to the existence of the edge, a smaller downshift is observed as shown in Fig. 3(f), in which the 2D peak is situated at $(2677 \pm 1)$ cm$^{-1}$.

In ref. 13, the 2D band is reported to downshift linearly with uniaxial strain in graphene. The amount of the downshift is ~ 20 cm$^{-1}$ for 1% strain, which seems to be much larger than our results for nonuniaxial strain. More experiments as well as theoretical investigations are needed for quantitative understanding of our results.

Finally, we comment on the possible damage of graphene due to e-beam exposure. Although we have not observed any enhancement of the D band in Raman spectra (not shown), which indicates the existence of defects in graphene, it is possible that excess e-beam exposure causes nucleation of defects in graphene.[14] We can eliminate this possibility by modifying the fabrication process; instead of exposing the substrate with graphene to e-beam, one can deposit graphene on LOR after the excess e-beam exposure by using the graphene transfer method common for CVD-grown graphene.[15] This modification will be indispensable for the application of our technique to graphene FETs.

In conclusion, we have developed a technique to introduce nonuniform strain in graphene by using pillars made of an electron beam resist. The amount of stretch, estimated from SEM images, reached ~ 20%. Raman spectroscopy indicated the application of nonuniform strain. This technique has a possibility to be applied to the formation of band gaps in graphene.

Acknowledgements

This work is partly supported by Grants-in-Aid for Scientific Research (22540329 and 23103503) from the Japan Society for the Promotion of Science. Raman spectroscopy was carried out in the Nano-Fabrication and Characterization Facility in NIMS. A. K. would like to thank A. K. Geim, H. Yoshioka, M. Hayashi, and K. Ishibashi for valuable comments.

References

1) A. H. C. Neto, F. Guinea, N. M. R. Peres, K. S. Novoselov, and A. K. Geim: Rev. Mod. Phys. **81** (2009) 109.

2) Y. W. Son, M. L. Cohen, and S. G. Louie: Phys. Rev. Lett. **97** (2006) 216803.

3) H. Min, B. Sahu, S. K. Banerjee, and A. H. MacDonald: Phys. Rev. B **75** (2007) 155115.

4) D. C. Elias, R. R. Nair, T. M. G. Mohiuddin, S. V. Morozov, P. Blake, M. P. Halsall, A. C. Ferrari, D. W. Boukhvalov, M. I. Katsnelson, A. K. Geim, and K. S. Novoselov: Science **323** (2009) 610.

5) S. Cheng, K. Zou, F. Okino, H. Rodriguez Gutierrez, A. Gupta, N. Shen, P. C. Eklund, J. O. Sofo, and J. Zhu: Phys. Rev. B **81** (2010) 205435.

6) M. Y. Han, J. C. Brant, and P. Kim: Phys. Rev. Lett. **104** (2010) 056801.

7) H. Miyazaki, K. Tsukagoshi, A. Kanda, M. Otani, and S. Okada: Nano Lett. **10** (2010) 3888.

8) F. Guinea, M. I. Katsnelson, and A. K. Geim: Nat. Phys. **6** (2010) 30.

9) N. Levy, S. A. Burke, K. L. Meaker, M. Panlasigui, A. Zettl, F. Guinea, A. H. Castro Neto, and M. F. Crommie: Science **329** (2010) 544.

10) K. S. Novoselov, A. K. Geim, S. V. Morozov, D. Jiang, Y. Zhang, S. V. Dubonos, I. V. Grigorieva, and A. A. Firsov: Science **306** (2004) 666.

11) A. C. Ferrari, J. C. Meyer, V. Scardaci, C. Casiraghi, M. Lazzeri, F. Mauri, S. Piscanec, D. Jiang, K. S. Novoselov, S. Roth, and A. K. Geim: Phys. Rev. Lett. **97** (2006) 187401.

12) D. Graf, F. Molitor, K. Ensslin, C. Stampfer, A. Jungen, C. Hierold, and L. Wirtz: Nano Lett. **7** (2007) 238.

13) Z. H. Ni, T. Yu, Y. H. Lu, Y. Y. Wang, Y. P. Feng, and Z. X. Shen: ACS Nano **2** (2008) 2301.

14) D. Teweldebrhan and A. A. Balandin: Appl. Phys. Lett. **94** (2009) 013101.

15) K. S. Kim, Y. Zhao, H. Jang, S. Y. Lee, J. M. Kim, K. S. Kim, J. H. Ahn , P. Kim, J. Choi, and B. H. Hong: Nature **457** (2009) 706.

Figure captions:

Fig.1.

Fabrication process of nanopillars. (a) Graphene is deposited on a Si/SiO$_2$ substrate coated with LOR resist. (b) Excess e-beam is exposed for the pillar pattern. In the LOR removal, graphene is suspended before being dipped into water (c) but attached to the substrate after it (d). (e) and (f) are optical and SEM images of suspended graphene with gold electrodes, respectively. The scale bar in (f) is 2.5 μm.

Fig. 2.

(a) SEM image of a graphene without lateral shift. The separation of adjacent pillars is 1 μm. (b)(c) Schematic side view of the sample before (b) and after (c) the LOR removal without the lateral shift of graphene. The graphene section between pillars stretches to attach to the substrate. From this model, one can estimate the average stretch between pillars. The result for image (a) is shown in (d). (e) SEM image of graphene with noticeable lateral shift. Dashed line and arrow indicate an edge of graphene and the shift direction, respectively.

Fig. 3.

Micro-Raman spectroscopy of strained graphene. (a) SEM image of the sample. SLG and MLG denote single layer and multilayer graphene, respectively. The Raman intensity between 2665 and 2675 cm$^{-1}$ for the area indicated by a yellow square is plotted in (b) in gray scale. Here, bright (dark) regions correspond to high (low) intensity. Red circles depict the pillars. (d)-(f) Raman spectra around 2D band are shown (d) for graphene on pillars, (e) for graphene near the midpoint of adjacent pillars, and (f) for graphene near an edge. Positions for each spectrum are indicated in (c). (g) Typical Raman spectrum for graphene without strain.

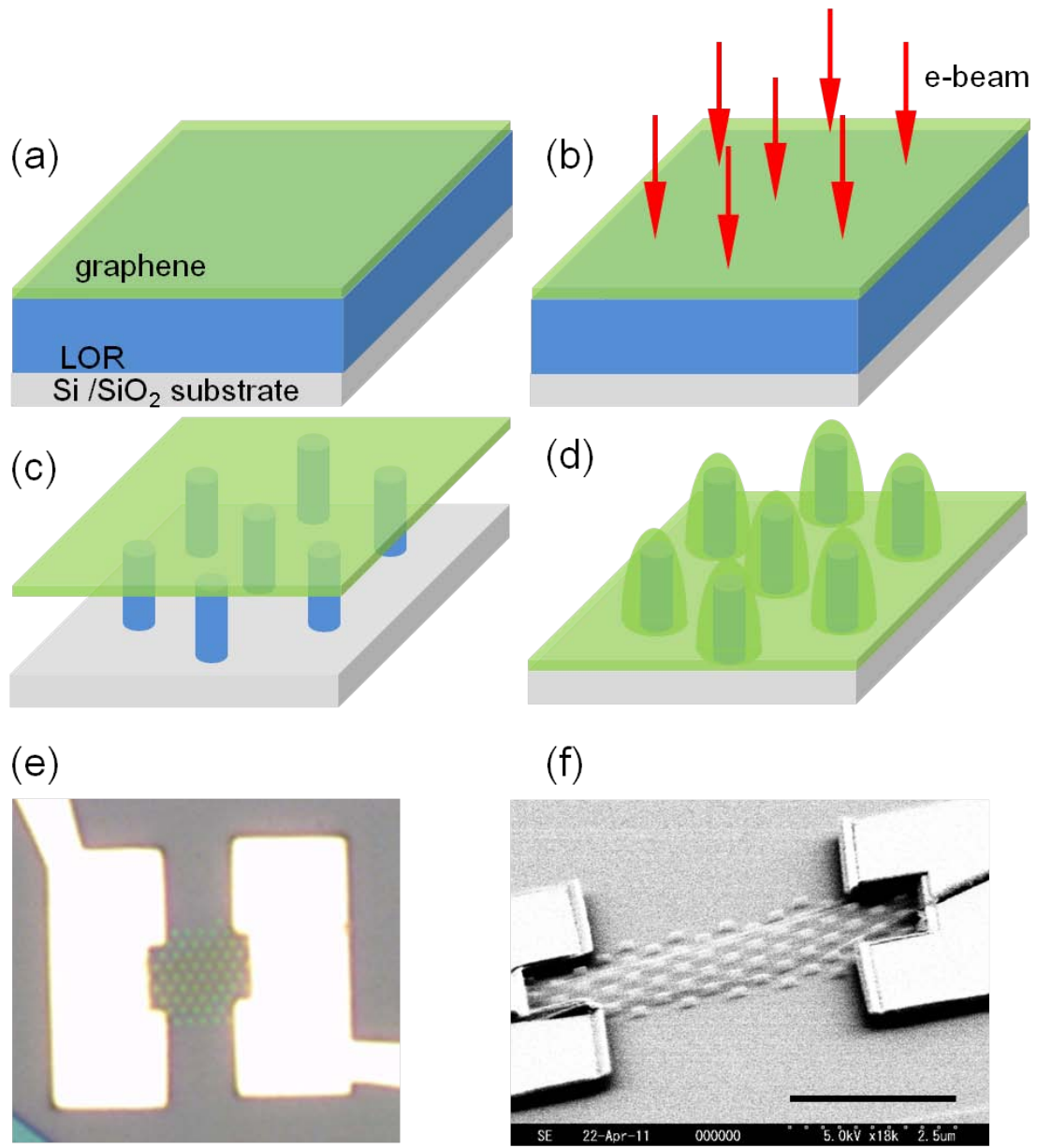

Fig. 1, H. Tomori et al.

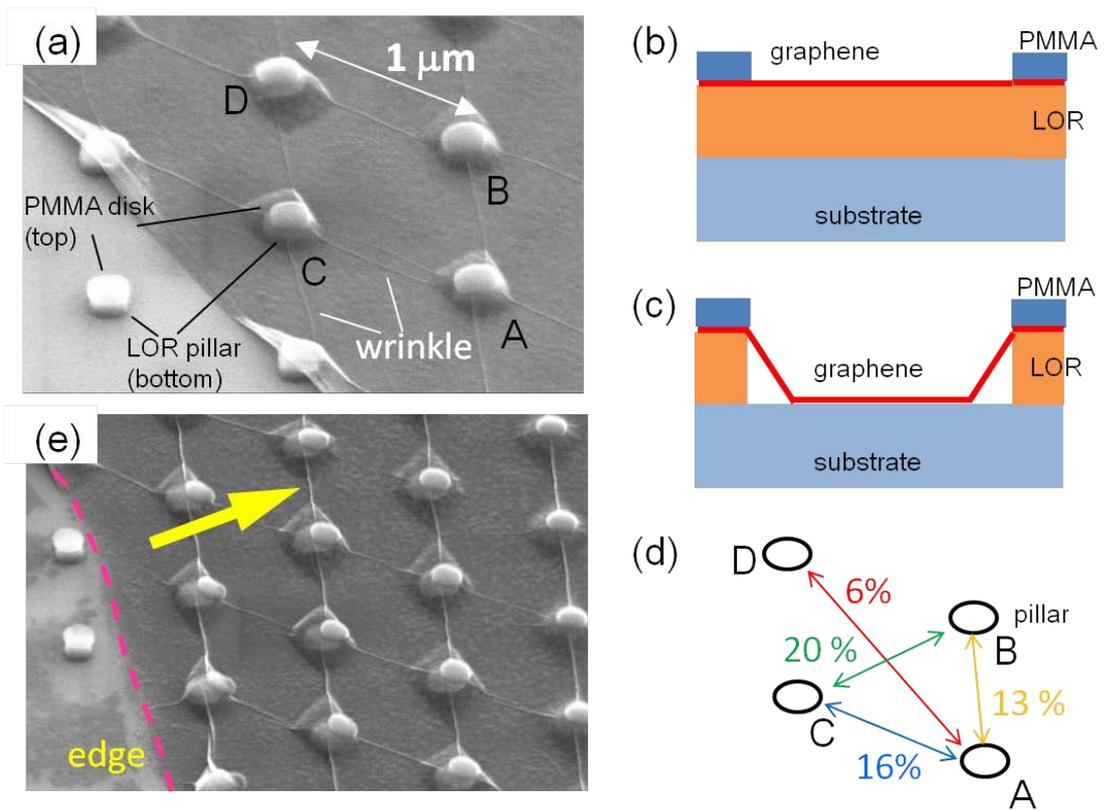

Fig. 2, H. Tomori et al.

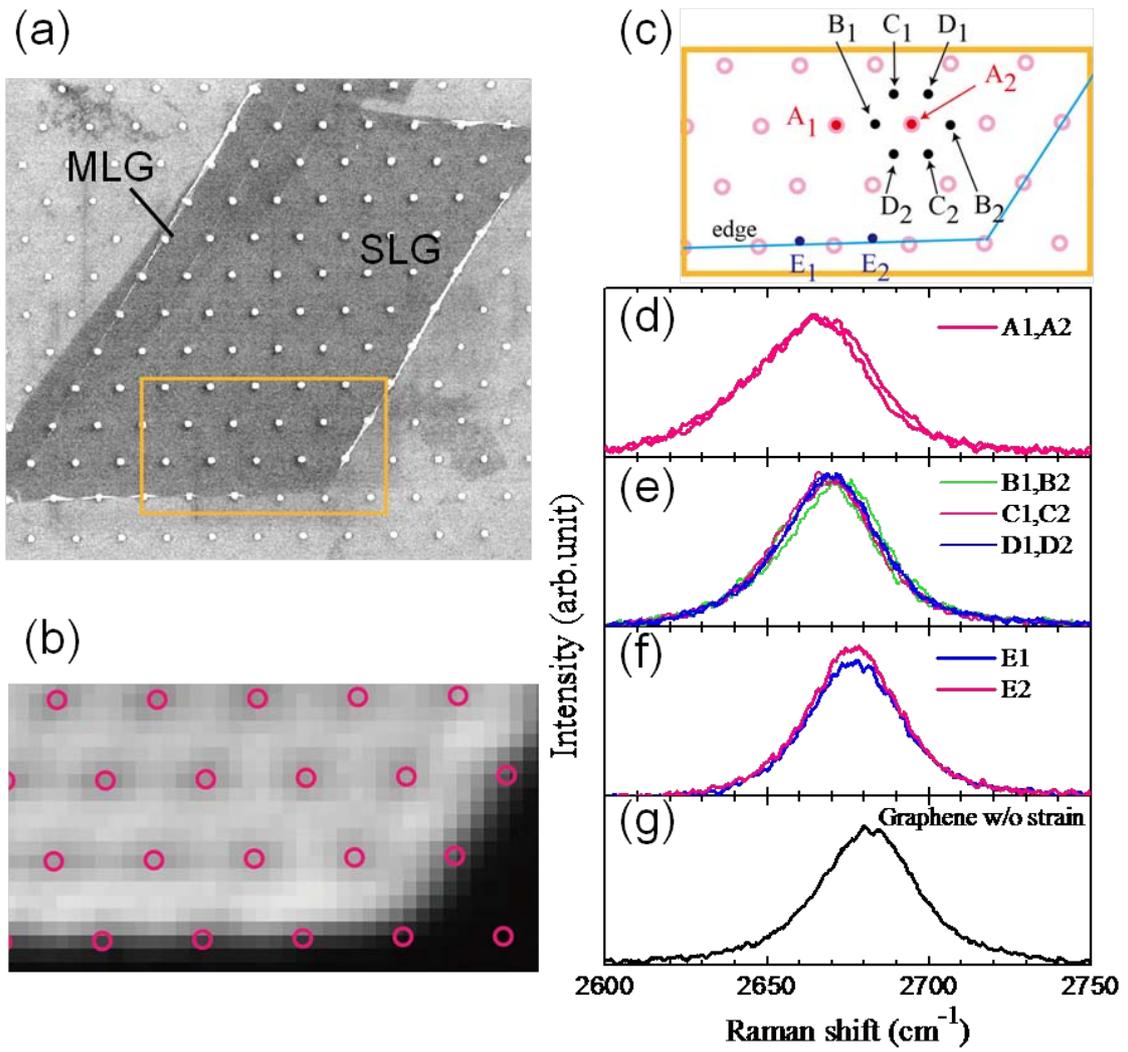

Fig. 3, H. Tomori et al.